\documentclass[aps,preprint,nofootinbib,superscriptaddress]{revtex4-1}
\pdfoutput=1
\usepackage{amsmath,slashed,amssymb,epsfig,amsthm,bm,graphicx}

\usepackage{hyperref}
\usepackage{color}
\definecolor{darkgreen}{rgb}{0,0.35,0}

\newcommand{\ucharles}{Faculty of Mathematics and Physics, Charles University, V Hole\v{s}ovi\v{c}k\'ach 2, 18000 Prague 8, Czech Republic}
\newcommand{\ubenha}{Physics Department, Faculty of Science, Benha University, Benha 13518, Egypt}

\newcommand{\unoka}{Irving K. Barber School of Arts and Sciences, University of British Columbia-Okanagan, Kelowna, BC V1V 1V7, Canada}
\newcommand{\unlet}{Deptartment of Physics and Astronomy, University of Lethbridge, Lethbridge, Alberta, T1K 3M4, Canada}
\newcommand{\kulak}{KU Leuven Campus Kortrijk - KULAK, Department of Physics, Etienne Sabbelaan 53, 8500 Kortrijk, Belgium}

\begin{document}

\title{Generalized Dirac structure beyond the linear regime in graphene}

\author{A.~Iorio}
\email{alfredo.iorio@mff.cuni.cz}
\affiliation{\ucharles}
\author{P.~Pais}
\email{pais@ipnp.troja.mff.cuni.cz}
\affiliation{\ucharles}
\affiliation{\kulak}
\author{I.~A.~Elmashad}
\email{ibrahim.elmashad@fsc.bu.edu.eg}
\affiliation{\ubenha}
\author{A.~F.~Ali}
\email{ahmed.ali@fsc.bu.edu.eg}
\affiliation{\ubenha}
\author{Mir~Faizal}
\email{mirfaizalmir@googlemail.com}
\affiliation{\unoka}
\affiliation{\unlet}
\author{L.~I.~Abou-Salem}
\email{loutfy.Abousalem@fsc.bu.edu.eg}
\affiliation{\ubenha}

\begin{abstract}
We show that a generalized Dirac structure survives beyond the linear regime of
the low-energy dispersion relations of graphene. A generalized uncertainty principle of the kind compatible with
specific quantum gravity scenarios with a fundamental minimal length (here graphene lattice spacing) and Lorentz violation (here the particle/hole asymmetry, the trigonal warping, etc.) is naturally obtained. We then show that the corresponding emergent field theory is a table-top realization of such scenarios, by explicitly computing the third order Hamiltonian, and giving the general recipe for any order. Remarkably, our results imply that going beyond the low-energy approximation does not spoil the well known correspondence with analogue massless quantum electrodynamics phenomena (as usually believed), but rather it is a way to obtain experimental signatures of quantum-gravity-like corrections to such phenomena.
\end{abstract}


\maketitle


\section{Introduction}

\noindent Graphene is popular among high energy theorists because, at small energies,
its electronic transport properties are in many respects those of a free gas
of massless particles, as theoretically discovered in Ref.~\cite{grapheneqft} and
experimentally proved in Ref.~\cite{geimnovoselovDiracElectrons}. Indeed, the dynamics of
the $\pi$ electrons of flat, unstrained, defects-free, monolayer graphene membranes,
for energies up to $\sim 3$eV above and below the Dirac point, is governed at a high level of accuracy by the Hamiltonian of a massless, Dirac, (2+1)-dimensional field, hence linear in the momentum, but propagating at $v_F \sim c / 300$ rather than at $c$ (on this see, e.g., the reviews Ref.~\cite{pacoreview2009} and Ref.~\cite{gusynin}).

It is important for high energy theory to have found here this special
relativistic-like behavior, as it gives us the opportunity to study elusive phenomena
through table top experiments. Nonetheless, what is more urgent is to test theories
\textit{beyond} special relativity, something difficult
(sometimes impossible) to do in direct experiments.
The use of graphene to this latter end was proposed in Ref.~\cite{weylgraphene},
and elaborated in follow-up works, see, e.g., Refs.~\cite{hawkinggrapheneplb,hawkinggrapheneprd,IorioPais,icrystals} and the
review Ref.~\cite{grapheneqftincurvedspace}.

Those works rely on the linear regime of the Dirac description,
hence are valid at small energies, and when curvature is considered,
the energies need to be even smaller\cite{grapheneqftincurvedspace}.
Here we move in the opposite direction, namely, we investigate what happens to the
Dirac structure in a range of energy beyond the $\pm 3$eV.
In fact, it is generally believed that already at the next step beyond the linear regime,
since the standard Dirac structure is gone, with it goes also the usefulness of graphene as a
lab for fundamental theories.

We show here that this is not the case, because a Dirac structure, although of a
generalized kind, is present beyond the linear approximation, and it is of the kind
obtained earlier in the research about generalized uncertainty principles\cite{DasVagenas,AhmedAli1,AhmedAli2,AhmedAli3} (GUPs) descending from quantum gravity
scenarios with a fundamental minimal length
(for a review see Ref.~\cite{Amelino-Camelia}).
These recently proposed GUPs differ considerably from other GUPs in the literature,
see, e.g. Ref.~\cite{Maggiore}, and the latter GUPs would not apply here.
The recent proposal is also consistent with doubly special relativity\cite{Amelino-Camelia} (DSR), where
there exists a maximum energy scale, that is the Planck
energy. At that scale the smooth manifold structure of spacetime breaks
down, along with the local Lorentz invariance. In graphene as well there is a maximum energy scale,
at which lattice effects break the smooth manifold description. This motivates our present investigation.

To evoke a concrete scenario where our present analysis could find a direct application, let us mention here
that almost all approaches to quantum gravity predict logarithmic corrections to the Bekenstein area law
for black hole entropy, see, e.g., Ref. \cite{DasMajumdarBhaduri}. Among those, the generalized uncertainty principle
also generates such correction terms\cite{AliLogArea}. On the other hand, it has been shown\cite{hawkinggrapheneplb,hawkinggrapheneprd}
that graphene, in the linear approximation, when it is given certain specific morphologies, can reproduce the Hawking/Unruh phenomenon corresponding to
a classical spacetime background. Such linear behavior is precisely what gets lost when discrete spacetime
(quantum gravity) effects take place. Therefore, applying the findings of the present work to that case would be
a step forward in measuring the effects of the corrections and compare with the theoretical predictions of the different models. Of course, the
full answer to those important questions, within this correspondence with graphene, will only come when the phononic excitations
of the graphene membrane are properly described within a properly constructed gravity correspondence, that is still not available.

\section{Beyond linear approximation and generalized Dirac}

\begin{figure}
\begin{center}
\includegraphics[width=0.8 \textwidth]{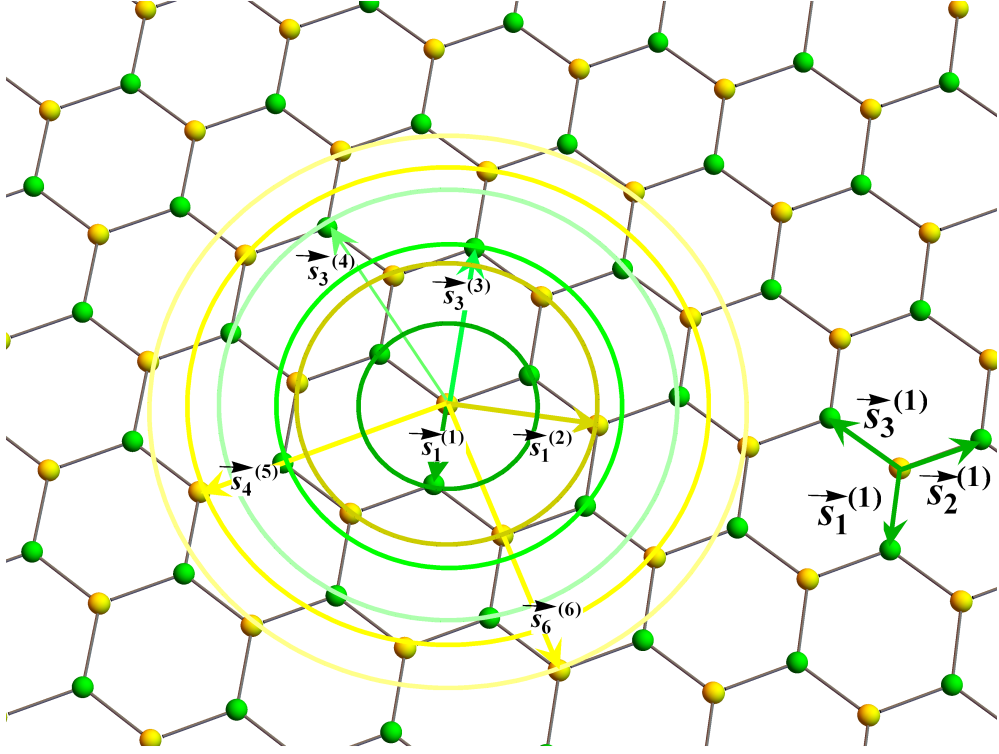}
\end{center}
\caption[3pt]{{\protect\small {The honeycomb lattice of carbons making graphene.
The two Bravais sublattices, $L_A$ and $L_B$, are identified by
different colors (green and yellow).
The three vectors connecting nearest neighbors are explicitly shown,
$\vec{s}^{(1)}_{1}= \ell(0,-1)$, $\vec{s}^{(1)}_{2}=\frac{\ell}{2}(\sqrt{3},1)$,
$\vec{s}^{(1)}_{3}=\frac{\ell}{2}(-\sqrt{3},1)$ with $\ell\approx1.42${\AA}
the carbon-to-carbon distance, whereas only one representative per $m$ of the
$\vec{s}^{(m)}_{i}$s, till $m=6$, is shown, along with the corresponding circle
encompassing all the relevant atoms. When the color of the circle is the same of
(different from) the color of the atom in the center, yellow here, the corresponding terms
contribute to the diagonal (off-diagonal) part of the Hamiltonian.
The less intense the color of the circle, the less important the associated terms (not is scale).
The $\vec{s}^{(m)}_{i}$s can be expressed in terms of the $\vec{s}^{(1)}_{i}$,
a shown in the Appendix~\ref{appendix_structure} for $m=2,...6$. }}}%
\label{honeycomb}%
\end{figure}

The honeycomb lattice in Fig.~\ref{honeycomb} has \textit{two} atoms per unit cell, what makes the lattice \textit{two} interpenetrating Bravais lattices. The quantum state associated to such configuration is
\begin{equation}
  |\Phi_{\vec{k}} \rangle =  a_{\vec{k}} |\phi^{A}_{\vec{k}} \rangle + b_{\vec{k}} |\phi^{B}_{\vec{k}} \rangle
\end{equation}
where ${\vec{k}}$ labels the crystal quasi-momentum, $a_{\vec{k}}$ and $b_{\vec{k}}$ are complex functions, and $\phi^{I}_{\vec{k}} (\vec{r}) = \langle \vec{r}|\phi^{I}_{\vec{k}} \rangle = e^{i \vec{k} \cdot \vec{r}} \varphi^{I} (\vec{r})$, with $I = A, B$, are Block functions, with periodic orbital wave-functions $\varphi^{I} (\vec{r})$ localized at the sites of the Bravais sublattices, either $L_A$ or $L_B$. That means that the shape of the functions $\varphi^{A}$ and $\varphi^{B}$ is the same, but the support is shifted. This localization is the key ingredient of the tight binding approximation: the more the $\varphi^{I} (\vec{r})$ are localized, the more accurate is the approximation (see, e.g., Ref.~\cite{altlandsimons}). Our goal now is to find the effective Hamiltonian that gives a good approximation of the graphene dispersion relations $E(\vec{k})$.

From the Schr\"{o}dinger equation $\hat{H}_{\vec{k}} |\Phi_{\vec{k}} \rangle = E(\vec{k}) |\Phi_{\vec{k}} \rangle$, multiplied from the left by $\langle \Phi_{\vec{k}}|$, and with
$\mathbf{1} = \int d^2 {\vec{r}}  |\vec{r} \rangle \langle {\vec{r}}| $ inserted, we have
\begin{eqnarray}
  {\cal H}_{\vec{k}} & \equiv & \langle \Phi_{\vec{k}}| \hat{H}_{\vec{k}} |\Phi_{\vec{k}} \rangle  \nonumber \\
   &=& \int d^2 {\vec{r}} d^2 {\vec{r}'} \langle \Phi_{\vec{k}} |\vec{r} \rangle \langle {\vec{r}}| \hat{H}_{\vec{k}} |\vec{r}' \rangle \langle {\vec{r}}' |\Phi_{\vec{k}} \rangle
   \label{fieldHfirst}
\end{eqnarray}
where ${\cal H}_{\vec{k}}$ is the ${\vec{k}}$ component of a ``field Hamiltonian'', that is the one that includes the wave functions. Now, we rewrite the overall interaction has sum of interactions between subsequent orders of neighbors, (including all orders this is merely a rewriting, with no approximation). Therefore, in (\ref{fieldHfirst}) we shall use $|\vec{r}' \rangle = |\vec{r} + \vec{s}^{(m)}_i \rangle$, where the vectors $\vec{s}^{(m)}_i$ join $m^{\textrm{th}}$-near neighbors, as indicated in Fig.~\ref{honeycomb}. With this the Hamiltonian splits into an onsite orbital part and an interaction potential $\langle {\vec{r}}| \hat{H}_{\vec{k}}  |\vec{r} + \vec{s}^{(m)}_i \rangle \rightarrow H^{(\textrm{o})}_{\vec{k}} (\vec{r}, \vec{r}) + V^{(m)}_{\vec{k}} (|\vec{r}' - \vec{r}|)$, where $\vec{r}' - \vec{r} = \vec{s}^{(m)}_i$.

The $\vec{s}^{(m)}_i$s, for any $m$ can be obtained from the $\vec{s}^{(1)}_{i}$s that we give in caption of Fig.1 and in the Appendix \ref{appendix_structure} (there along with $\vec{s}^{(m)}_i$s for $m =2,...,6$). $V^{(m)}_{\vec{k}}$ is the potential between atoms located at distance $|\vec{s}^{(m)}_i| \equiv |\vec{s}^{(m)}|$, $\forall i = 1,..., n_m$ (in the ideal case of a perfect lattice), while $H^{(\textrm{o})}_{\vec{k}}$ is the Hamiltonian whose eigenfunctions are the orbitals $\varphi^{I} (\vec{r})$ with corresponding eigenvalues $\epsilon^{(0)}$ that is the onsite energy.

Putting together all this, the field Hamiltonian (\ref{fieldHfirst}) becomes
\begin{eqnarray}
  {\cal H}_{\vec{k}} & = & \sum_{m \in \textrm{diag}} (\epsilon^{(0)}  \varsigma_m + \eta_m )  {\cal F}_m ({\vec{k}}) (a^*_{\vec{k}} a_{\vec{k}} + b^*_{\vec{k}} b_{\vec{k}}) \nonumber \\
  &+& \left( \sum_{m \in \textrm{off}} (\epsilon^{(0)}  \varsigma_m + \eta_m )  {\cal F}^*_m ({\vec{k}}) a^*_{\vec{k}} b_{\vec{k}} + h.c. \right)   \label{fieldHthird}
\end{eqnarray}
where $\varsigma_m = \int d^2 \vec{r} \varphi^{* I} (\vec{r}) \varphi^J (\vec{r} + \vec{s}^{(m)}_i)$ and
$\eta_m = \int d^2 \vec{r} \varphi^{* I} (\vec{r}) V^{(m)} \varphi^J (\vec{r} + \vec{s}^{(m)}_i)$ are the overlapping and the hopping parameters, respectively, and the $m$s contribute to the diagonal terms when $I = J$, and to the off-diagonal terms when $I \neq J$. Finally ${\cal F}_m ({\vec{k}}) \equiv \sum_{i=1}^{n_m} e^{i {\vec{k}} \cdot {\vec{s}^{(m)}}_i}$, where
\begin{equation}\label{firstfunction}
  {\cal F}_1 = \sum_{i= 1}^{3} e^{i {\vec{k}} \cdot {\vec{s}^{(1)}}_i} =
  e^{- i \ell k_y} [1 + 2 e^{i \frac{3}{2} \ell k_y} \cos(\frac{\sqrt{3}}{2} \ell k_x)]
\end{equation}
and ${\cal F}_2 = |{\cal F}_1|^2 - 3$. More details on the derivation of (\ref{fieldHthird}) and on the ${\cal F}_m$s are in the Appendix \ref{appendix_structure}.

By setting the zero of $E(\vec{k})$ by $\epsilon^{(0)}$ the Schr\"{o}dinger equation becomes (see Appendix~\ref{appendix_structure})
\begin{equation}\label{Shcroed3}
\psi^\dag_{\vec{k}} \widehat{{\cal H}}_{\vec{k}} \psi_{\vec{k}} = E(\vec{k}) \psi^\dag_{\vec{k}} \widehat{{\cal S}}_{\vec{k}} \psi_{\vec{k}}
\end{equation}
where $\psi^\dag_{\vec{k}} \equiv (a^*_{\vec{k}}, b^*_{\vec{k}})$ and
\begin{equation}\label{overlapmatrix}
\widehat{ {\cal H}}_{\vec{k}} =
  \left(
                        \begin{array}{cc}
                          h_{\vec{k}}^{\textrm{diag}} &  h^{* \textrm{off}}_{\vec{k}} \\
                           h^{ \textrm{off}}_{\vec{k}} &  h_{\vec{k}}^{\textrm{diag}} \\
                        \end{array}
                       \right)
                                          \; {\textrm{and}} \;
\widehat{ {\cal S}}_{\vec{k}} =  \left(
                        \begin{array}{cc}
                          {\cal S}_{\vec{k}}^{\textrm{diag}} &  {\cal S}^{* \textrm{off}}_{\vec{k}} \\
                           {\cal S}^{ \textrm{off}}_{\vec{k}} &  {\cal S}_{\vec{k}}^{\textrm{diag}} \\
                        \end{array}
                      \right)
\end{equation}
are the Hamiltonian and the overlapping matrices respectively, with
\begin{equation*}
  h_{\vec{k}}^{\textrm{diag}} \left( {\cal S}_{\vec{k}}^{\textrm{diag}} \right) =
 \sum_{m \in \textrm{diag}} \eta_m (\varsigma_m) {\cal F}_m ({\vec{k}}) \;,
\end{equation*}
and
\begin{equation*}
  h_{\vec{k}}^{\textrm{off}} \left( {\cal S}_{\vec{k}}^{\textrm{off}} \right) =
 \sum_{m \in \textrm{off}} \eta_m (\varsigma_m) {\cal F}_m ({\vec{k}}) \;.
\end{equation*}
Notice that $\widehat{{\cal H}}_{\vec{k}} = h_{\vec{k}}^{\textrm{diag}} \mathbf{{1}}_2 + \textrm{Re} h_{\vec{k}}^{\textrm{off}} \sigma_x
                      + \textrm{Im} h_{\vec{k}}^{\textrm{off}} \sigma_y$, with $\mathbf{{1}}_2 = \left(
                   \begin{array}{cc}
                   1 & 0 \\
                   0 & 1 \\
                   \end{array}
                 \right)
$,
$\sigma_x = \left(
                   \begin{array}{cc}
                     0 & 1 \\
                     1 & 0 \\
                   \end{array}
                 \right)
$, $\sigma_y = \left(
                   \begin{array}{cc}
                     0 & -i \\
                     i & 0 \\
                   \end{array}
                 \right)
$. Clearly the concurrence of a) two atoms per unit cell (doublet structure) and of b) the Hermiticity of the Hamiltonian are behind the emergence of the Pauli matrices. Nonetheless this is not enough to have a Dirac structure, as we need to focus on the dispersion relations descending from (\ref{Shcroed3}), that is the two solutions to the secular equation
${\rm det}   \left( \widehat{{\cal H}}_{\vec{k}} - E  \widehat{{\cal S}}_{\vec{k}} \right) = 0$ that gives an involved expression in terms of the ${\cal F}_m$s, and in general there is no Dirac structure behind such dispersion relations. In fact, the values of both, the overlapping and the hopping parameters, $\varsigma_m$ and $\eta_m$, respectively, exponentially drop with $m$, as should be expected. Although there is no absolute consensus on the actual values of these phenomenological parameters, as their definition might differ depending on the model, there is general agreement on the fact that, in the ideal case of an infinite and non-deformed sheet of graphene, the values of the first few of such parameters are (see, e.g., Ref.~\cite{MauroFerreira} and references therein)
\begin{eqnarray}
 \eta_0 &\simeq& - 0.36\textrm{eV},  \eta_1 \simeq - 2.8\textrm{eV} , \eta_2 \simeq 0.12\textrm{eV} , \nonumber \\
 \varsigma_1 &\simeq& 0.106 , \varsigma_2 \simeq 0.001
\end{eqnarray}
from which we see that stopping at $\eta_2$ and $\varsigma_1$ is a very good approximation. For the sake of an easier comparison with the literature, it is worth mentioning here that $\eta_1$ is what is usually indicated with $t$, while $\eta_2$ is what is usually indicated with $t'$, see, e.g., Ref.\cite{pacoreview2009}.

Once this happens, the secular equation becomes
\begin{eqnarray}\label{fullDispOrd2}
E_\pm & = & \left( 1 - \varsigma_1^2 |{\cal F}_1|^2 \right)^{-1}  \\
& \times & \left(\epsilon_0 - \epsilon_1 |{\cal F}_1|^2 \pm \sqrt{(\epsilon_0 - \epsilon_1 |{\cal F}_1|^2)^2
- ( 1 - \varsigma_1^2 |{\cal F}_1|^2) [(\epsilon_0 + \eta_2 |{\cal F}_1|^2)^2 - \eta^2_1 |{\cal F}_1|^2 ]} \right)  \nonumber
\end{eqnarray}
where $\epsilon_0 \equiv \eta_0 - 3 \eta_2$ and $\epsilon_1 \equiv \varsigma_1 \eta_1 - \eta_2$. The fact that, based on the result ${\cal F}_2 = |{\cal F}_1|^2 - 3$ (see Appendix \ref{appendix_structure}), in (\ref{fullDispOrd2}) we can expand $E_\pm$ in terms of one function, ${\cal F}_1$, is what shall give us the possibility to identify a generalized Dirac structure of the kind obtained in Refs.~\cite{AhmedAli1,AhmedAli2,AhmedAli3}. This occurrence, though, stops at the second step beyond nearest neighbors, because  ${\cal F}_3 \nsim  {\cal F}_1^3$, ${\cal F}_4 \nsim  {\cal F}_1^4$, etc. (see Appendix \ref{appendix_structure}). Therefore, although some form of recursion is clearly present, e.g., ${\cal F}_6 = |{\cal F}_3|^2 - 3$ (see Appendix~\ref{appendix_structure}), that does not give a straightforward power expansion in terms of ${\cal F}_1$. This deserves further study.

At the Dirac points\footnote{In this paper, since no defects, nor deformations of the membrane are considered, we can safely focus on a single Dirac point, say $\vec{K}_{1}\equiv\vec{K}^{D}$. This we do from now on.}, i.e. the solutions of ${\cal F}_1 (\vec{k}) = 0$, $\vec{K}_{1} = \left(\frac{4 \pi}{3 \sqrt{3} \ell}, 0 \right)$ and $\vec{K}_{2} = \left(-\frac{4 \pi}{3 \sqrt{3} \ell}, 0 \right)$, we have $E_\pm|_{{\cal F}_1 = 0} = \epsilon_0$, that is the shift of the zero of the energy due to diagonal terms in the Hamiltonian. We can shift once more the zero of the energy, hence dropping this $\epsilon_0$ but shifting all other values accordingly, in particular $\epsilon_1 \to {\epsilon'}_1 \equiv \epsilon_1 - \epsilon_0$. Thus, finally, dropping higher order terms $O(\varsigma_1^2)$ and $O(\varsigma_1 \eta_2)$, we have
\begin{equation} \label{DispRelWithFunct}
  E_\pm = \eta_1 \left( \pm |{\cal F}_1| - \tilde{A}  |{\cal F}_1|^2 \right)
\end{equation}
where $\tilde{A} = {\epsilon'}_1/ \eta_1$ is a dimensionless parameter. If we define the dimensionless vector ${\vec{\tilde P}} \equiv (Re {\cal F}_1, Im {\cal F}_1)$ then
\begin{eqnarray}
  E_\pm & = & \eta_1 \left[ \pm \frac{\ell}{\hbar} \; \frac{\hbar}{\ell} |{\vec{\tilde P}}| - \tilde{A} \; \left( \frac{\ell}{\hbar} \; \frac{\hbar}{\ell} \right)^2 |{\vec{\tilde P}}|^2 \right] \\
  & \equiv & V_F \left( \pm |\vec{P}| - A \; |\vec{P}|^2 \right) \label{DispRelLargeP}
\end{eqnarray}
where ${\vec{P}} \equiv(\hbar / \ell) {\vec{\tilde P}}$, $A \equiv (\ell / \hbar ) \tilde{A}$, $V_F \equiv \eta_1 \ell / \hbar$ are the dimension-full quantities (recall that\cite{pacoreview2009, gusynin} $v_F \simeq 1.5 \eta_1 \ell / \hbar$).

Henceforth, considering ${\vec{P}}$ as a {\it generalized momentum}, we can use the customary Dirac prescription $|{\vec{P}}| \rightarrow \vec{\sigma} \cdot {\vec{P}}$ that gives
\begin{equation}\label{newDiracHwithP}
  H (P) \equiv \sum_{\vec{k}} {\cal H}_{\vec{k}} =
  V_F  \sum_{\vec{k}} \psi^\dag_{\vec{k}} \left(\not\!P - A \; \not\!P \not\!P \right) \psi_{\vec{k}}
\end{equation}
as the effective Hamiltonian corresponding to the dispersion relations (\ref{DispRelLargeP}). Here $\not\!P \equiv \vec{\sigma} \cdot {\vec{P}}$ and we used the properties of the Pauli matrices, $\sigma_i \sigma_j = \delta_{i j} \mathbf{1}_2 + i \epsilon_{i j k} \sigma^k $, that give $\not\!P \not\!P =  |\vec{P}|^2 \mathbf{1}_2$. This is precisely the form of the Hamiltonian obtained in Refs.~\cite{AhmedAli1,AhmedAli2,AhmedAli3} when generalizing the Dirac Hamiltonian to allow for a GUP with a minimal fundamental length, here given by the $\ell$.  Now we can proceed according to two options, both giving the same deformed/generalized Dirac equation (hence, the same dispersion relations (\ref{DispRelLargeP})).

The first option consists in taking the \textit{deformed} Hamiltonian (\ref{newDiracHwithP}), together with \textit{standard} commutation relations, $[X_i^P, P_j] = i \hbar \delta_{i j}$, where $X_i^P$ are generalized conjugate variables of the generalized momenta $P_i$.

The second option consists in introducing a new generalized momentum
\begin{equation}\label{Q}
{\vec{Q}} \equiv {\vec{P}} (1 - A |{\vec{P}}|)
\end{equation}
of which the former, $\vec{P}$, is the low-energy approximation, in terms of which we have a \textit{standard} Dirac Hamiltonian\footnote{This needs be done together with the previously used customary Dirac prescription, $|{\vec{P}}| \rightarrow \vec{\sigma} \cdot {\vec{P}}$.}
\begin{equation}\label{newDiracHwithQ}
H (Q) = V_F \sum_{\vec{k}} \psi^\dag_{\vec{k}} \not\!Q \psi_{\vec{k}}
\end{equation}
This time, though, the Weyl-Heisenberg algebra is \textit{deformed} to
\begin{equation}\label{QGUP}
\left[X^P_{i},Q_{j}\right] = i\hbar \left[ \delta_{ij} - A \left( Q \delta_{ij} + \frac{Q_i Q_j}{Q} \right)\right]
\end{equation}
that is precisely the GUP of \cite{AhmedAli1,AhmedAli2,AhmedAli3}, when coordinates commute among themselves, and so do momenta (see especially \cite{AhmedAli3}).

In both the equivalent cases, we have effective Hamiltonians that govern the physics of graphene within a range of energies much wider than the one where the linear approximation holds.

\section{Modified Weyl-Heisenberg algebra and Lorentz symmetry breaking}

We have here three momenta, two generalized momenta, $\vec{Q}$ and $\vec{P}$, and one low-energy standard momentum $\vec{p}$. Notice that $\vec{P}$ is low-energy with respect to $\vec{Q}$ (next-to-nearest neighbors vs nearest neighbors), but it is high energy with respect to $\vec{p}$, that is the only standard low-energy momentum arising from expanding $\vec{P}$, therefore ${\cal F}_1 ({\vec{k}})$ in (\ref{firstfunction}), around the Dirac point ${\vec{k}} = {\vec{K}^D} + {\vec{p}}$. In fact, we must do that since the structure in (\ref{newDiracHwithP}) is only apparently $O(\ell)$, because ${\vec{P}}$, as it stands, contains all orders in $\ell$. Before doing so, though, let us comment on the algebraic structures that we have obtained.

Each momentum has its conjugate variable, $[X_i^Q, Q_j] = i \hbar \delta_{i j}$, $[X_i^P, P_j] = i \hbar \delta_{i j}$, and $[x_i, p_j] = i \hbar \delta_{i j}$, but when we compute, e.g., commutators between low-energy coordinates and high-energy momenta, the standard Weyl-Heisenberg algebra is modified, as in (\ref{QGUP}). Nonetheless, the low-energy coordinates are $x_i$, hence to see the effects of the corrections we must ultimately compute $[x_i, Q_j]$. We shall give those commutators, $[x_i, Q_j]$s,  in a moment, before that, let us comment on the important issue of Lorentz symmetry.

At the first level of the ladder, $(X_i^Q, Q_j)$, full Lorentz symmetry is preserved. At the second level, $(X_i^P, Q_j)$, the Hamiltonian (\ref{newDiracHwithP}) preserves rotation symmetry, $SO(2)$ in this case, but full Lorentz is gone, e.g., the particle/antiparticle symmetry (particle/hole), as it is clear from (\ref{DispRelLargeP}) and from the terms $\frac{Q_i Q_j}{Q}$ in (\ref{QGUP}). At the last level in the ladder, $(x_i, Q_j)$, not even rotation symmetry of the Hamiltonian survives, and we expect rotation-symmetry violating terms. Indeed, in the standard treatment such terms are seen (see, e.g, the trigonal warping\cite{pacoreview2009}), and we naturally reproduce them in this new perspective. This is a rich algebraic structure, worth investigating, especially in relation to the possibility for noncommutativity.

When we expand $\vec{P} (\vec{p})$, we obtain the following dispersion relations, up to $O(p^3)$ (see (\ref{DispRelLargeP}))
\begin{eqnarray}\label{DispRelsmallp}
  E_\pm  & = & V_F \big( \pm \frac{3}{2} p \mp  \frac{3}{8} \ell p^2 \cos(3\theta) \mp  \frac{3}{64} \ell^2 p^3 \cos^2(3\theta)   \nonumber \\
         & - & A \frac{9}{8} p^2 + A \frac{9}{8} \ell p^3 \cos(3\theta) \big)   \nonumber \\
         & = & v_F \big( \pm p \mp  \frac{\ell}{4} p^2 \cos(3\theta) \mp  \frac{\ell^2}{32}  p^3 \cos^2(3\theta)   \nonumber \\
         & - & \frac{3 \ell}{4} \frac{{\epsilon'}_1}{\eta_1} p^2 + \frac{3 \ell^2}{4} \frac{{\epsilon'}_1}{\eta_1} p^3 \cos(3\theta) \big) \;,
\end{eqnarray}
where $p = |\vec{p}|$, $\tan \theta = p_y / p_x$, $v_F \equiv 3/2 V_F$, and $\hbar = 1$.

\begin{figure}
\begin{center}
\includegraphics[width=.8\textwidth]{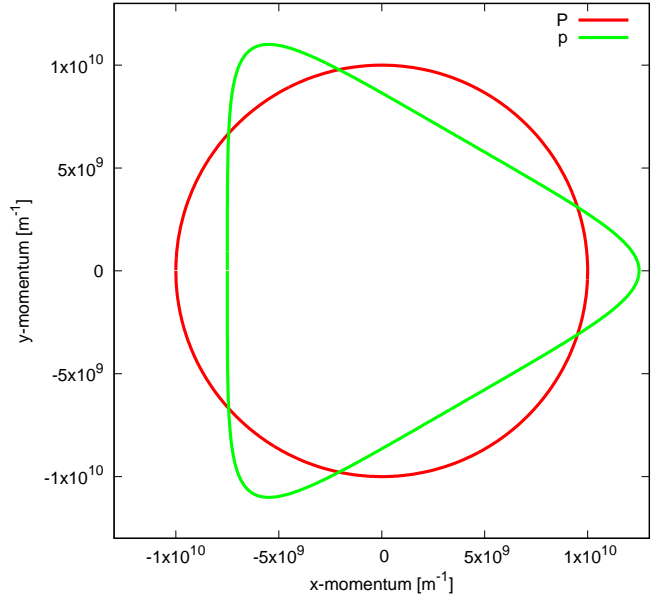}
\caption{In order to see clearly the non-isotropic effect of the transformation, we plot the mapping $P \to p$ given in \eqref{high-low_transformation} acting on a circle of radius $1.0\times10^{-10}$m$^{-1}$ in the momenta space up to order $O(\ell)$, where $\ell=1{\AA}$.}
\label{low-high_transf_fig}
\end{center}
\end{figure}

Lorentz violating terms appear in the dispersion relation \eqref{DispRelsmallp}. As customary in Lorentz violating scenarios (see, e.g, Refs.~\cite{Iorio2007,Iorio2008}), the orientation of the vectors is important. Here all vectors will be referred to the honeycomb lattice orientation, and we must mention explicitly our axis choice, see Fig.~\ref{honeycomb}.

Accordingly the corresponding $O(p^3)$ Hamiltonian is
\begin{eqnarray}\label{HamiltonianOp3}
H & = & v_F \sum_{\vec{p}} \psi^\dag_{\vec{p}} \big[ \sigma_x \left( p_x - \frac{\ell}{4} (p^2_x - p^2_y) - \frac{\ell^2}{8} p_x (p^2_x + p^2_y) \right) \nonumber \\
  & + & \sigma_y \left( p_y + \frac{\ell}{2} p_x p_y - \frac{\ell^2}{8} p_y (p^2_x + p^2_y) \right) \nonumber \\
  & - & \frac{3}{2} \frac{{\epsilon'}_1}{|\eta_1|} \left( \ell (p^2_x + p^2_y) - \frac{\ell^2}{2} p_x^3 + \frac{3 \ell^2}{2} p_x p^2_y \right) \big]  \psi_{\vec{p}}
\end{eqnarray}
whose structure would hardly induce to think of any Dirac structure, although, as shown in the previous part, in fact it is just a low energy realization of a generalized Dirac structure. Let us express the expansion of $\vec{P}$ in terms of $\vec{p}$ as follows $\vec{P} = \vec{p}+\Delta\vec{p}$, where
\begin{eqnarray}\label{high-low_transformation}
\Delta\vec{p}&=&\frac{\ell}{4}\left(p_{y}^{2}-p_{x}^{2}-\frac{\ell}{2}p_{x}(p^{2}_{x}+p^{2}_{y}),2p_{x}p_{y}-\frac{\ell}{2}p_{y}(p^{2}_{x}+p^{2}_{y})\right) \nonumber\\
&=&\frac{\ell}{4}p^{2}(-\cos2\theta-\frac{\ell}{2}p\cos\theta,\sin2\theta-\frac{\ell}{2}p\sin\theta)\;,
\end{eqnarray}
We can visualize this as transformation acting on the whole momenta plane, as plotted in Fig.~\ref{low-high_transf_fig} where we show the mapping $\vec{P}\to\vec{p}$ for a circle of radius $1.0\times10^{-10}$ m$^{-1}$ ($\ell=1{\AA}$).
Of course, in order to obtain the operator expression for $H$ in \eqref{HamiltonianOp3}, one needs to perform the customary substitution $p_i \to - i \partial/\partial x_i$.

Finally, after the expansion, the commutation relations one has to write are obtained by first noticing that now $\left[x_{i},P_{j}\right]\neq i\delta_{ij}$, because the role of the low energy momentum is now played by $\vec{p}$ and not by $\vec{P}$ ($\left[x_{i},p_{j}\right]=i\delta_{ij}$). Then, with the help of $\left[x_{i},P_{j}\right]$ and using \eqref{Q}, the commutations of the system are
$\left[x_{i},Q_{j}\right]$ fully expressed in terms of $Q_{i}$ (see details in Appendix \ref{appendix_commutators}). The actual structure changes according to the order $O(\ell^{n})$. For instance, to $O(\ell)$,
\begin{eqnarray}\label{modified_commutator}
\left[x,Q_{x}\right] & = & i \left(1 - \frac{\ell}{2} Q_{x}-\frac{\ell{\epsilon'}_1}{ \eta_1}\left(|\vec{Q}|+\frac{Q_{x}^{2}}{|\vec{Q}|}\right)\right) \; , \nonumber \\
\left[x,Q_{y}\right] & = & i \left( \frac{\ell}{2} Q_{y}-\frac{\ell{\epsilon'}_1}{ \eta_1|\vec{Q}|}Q_{x}Q_{y}\right)\; , \nonumber  \\
\left[y,Q_{x}\right] & = &  i \left( \frac{\ell}{2} Q_{y}-\frac{\ell{\epsilon'}_1}{ \eta_1|\vec{Q}|}Q_{x}Q_{y}\right)   \; , \\
\left[y,Q_{y}\right] & = & \left(1 + \frac{\ell}{2} Q_{x}-\frac{\ell{\epsilon'}_1}{ \eta_1}\left(|\vec{Q}|+\frac{Q_{y}^{2}}{|\vec{Q}|}\right)\right)\;. \nonumber
\end{eqnarray}
In the Appendix \ref{appendix_commutators} we give the result up to order $O(\ell^{2})$.

\section{Conclusions}

Our results imply that, contrarily to what customarily believed, going beyond the low-energy approximation does not spoil the well known correspondence with analogue massless quantum electrodynamics phenomena, such as the Hawking-Unruh phenomenon\cite{hawkinggrapheneplb,hawkinggrapheneprd} or  the Klein tunneling\cite{Geim}, to cite two. Rather, from here it is clear that one should consider high energy contributions as a way to obtain precious experimental signatures of quantum-gravity-like corrections to such phenomena. The quantum gravity scenarios here are those for which continuous
fields and continuous spacetimes are both emerging from a single granular underlying structure (see, e.g., Ref.~\cite{Kleinert1987} and also Ref.~\cite{iorioscholtzarXiv2017}).
Indeed, we are able to naturally reproduce the GUP of Refs.~\cite{AhmedAli1,AhmedAli2,AhmedAli3}, hence the quantum gravity scenario with a
fundamental scale\cite{Amelino-Camelia}. The phenomenology includes the particle/hole asymmetry
corresponding to matter/anti-matter asymmetry, and the non-isotropic trigonal warping terms
related to Lorentz violation, similar to those of Kostelecky Standard Model Extension\cite{kosteleckysamuel,colladaykostelecky1,colladaykostelecky2} (SME). The anisotropy of the
continuous field theory for graphene was seen in Ref.~\cite{IorioPais}, and anisotropy
of the universe has been observed, and called\cite{Magueijo,PlanckCollaboration} ``the axis of evil''.

Let us close by mentioning some directions that could be pursued from here. First, as mentioned in the Introduction, the immediate follow-up of this work is to perform the calculations of Refs.\cite{hawkinggrapheneplb,hawkinggrapheneprd} with the new Hamiltonian proposed here, that include lattice/granular spacetime effects. In that context, one could then compute the entanglement entropy
of the quantum fields in such nontrivial background with horizons, and compare with the quantum-gravity-corrected Bekenstein results. A second direction is based on the fact that other materials exhibit emergent pseudo-relativistic fermions\cite{Wang,Claudia}, hence, it would be interesting to perform a similar study for them. A further direction is a deeper study of the obtained algebraic structures, especially the exploration of the possibility that noncommuting (generalized) coordinates\cite{Doplicher1995, Doplicher2015} might be realized.

\section*{Acknowledgments}

A.~I. and P.~P acknowledge partial financial support from the Czech Science Foundation (GA\v{C}R), under contract no. 14-07983S.
P. P. is supported by a PDM grant of KU Leuven and was partially supported by Fondecyt Grant 1140155 during the first stage of this work.

\appendix


\section{Structure of the $m$-nearest neighbor Hamiltonian}
\label{appendix_structure}

In this Appendix we give some more details of the construction leading to the tight binding Hamiltonian in general, focussing on the structure

Let us start by listing the $\vec{s}^{(m)}_{i}$s vectors up to $m = 6$. First, the onsite $\vec{r}' = \vec{r}$ vector is clearly zero, $\vec{s}^{(0)} = 0$. Then, the nearest neighbor vectors, $ \vec{s}^{(1)}_{i}$, $i = 1,2,3$, are
\begin{equation}\label{sNN}
  \vec{s}^{(1)}_{1} = \ell(0,-1) \; , \vec{s}^{(1)}_{2} = \frac{\ell}{2}(\sqrt{3},1) \; , \vec{s}^{(1)}_{3}=\frac{\ell}{2}(-\sqrt{3},1)
\end{equation}
with $\ell\approx1.42${\AA} the carbon-to-carbon distance. The next to nearest neighbor vectors, $\vec{s}^{(2)}_{i}$, $i = 1,..., 6$, are
\begin{eqnarray}
  \vec{s}^{(2)}_{1} & = & \vec{s}^{(1)}_{2} - \vec{s}^{(1)}_{3} ,
  \vec{s}^{(2)}_{2} = \vec{s}^{(1)}_{2} - \vec{s}^{(1)}_{1} ,
  \vec{s}^{(2)}_{3} = \vec{s}^{(1)}_{3} - \vec{s}^{(1)}_{1} , \nonumber \\
  \vec{s}^{(2)}_{4} & = & - \vec{s}^{(2)}_{1} ,
  \vec{s}^{(2)}_{5} = - \vec{s}^{(2)}_{2} ,
  \vec{s}^{(2)}_{6} = - \vec{s}^{(2)}_{3} . \label{s2}
\end{eqnarray}
The $3^{rd}$-nearest neighbor vectors, $\vec{s}^{(3)}_{i}$, $i = 1,2,3$, are
\begin{eqnarray}
  \vec{s}^{(3)}_{1} = \vec{s}^{(2)}_{1} + \vec{s}^{(1)}_{1} ,
  \vec{s}^{(3)}_{2} = - \vec{s}^{(2)}_{2} + \vec{s}^{(1)}_{3} ,
  \vec{s}^{(3)}_{3} = \vec{s}^{(2)}_{3} - \vec{s}^{(1)}_{2} . \label{s3}
\end{eqnarray}
The $4^{th}$-nearest neighbor vectors, $\vec{s}^{(4)}_{i}$, $i = 1,...,6$, are
\begin{eqnarray}
  \vec{s}^{(4)}_{1} & = & \vec{s}^{(2)}_{1} + \vec{s}^{(1)}_{2} ,
  \vec{s}^{(4)}_{2} = \vec{s}^{(2)}_{2} + \vec{s}^{(1)}_{2} ,
  \vec{s}^{(4)}_{3} = \vec{s}^{(2)}_{3} + \vec{s}^{(1)}_{3} , \nonumber \\
  \vec{s}^{(4)}_{4} & = & - \vec{s}^{(2)}_{1} + \vec{s}^{(1)}_{3} ,
  \vec{s}^{(4)}_{5} = - \vec{s}^{(2)}_{2} + \vec{s}^{(1)}_{1}  ,
  \vec{s}^{(4)}_{6} = - \vec{s}^{(2)}_{3} + \vec{s}^{(1)}_{1} . \label{s4}
\end{eqnarray}
The $5^{th}$-nearest neighbor vectors, $\vec{s}^{(5)}_{i}$, $i = 1,..., 6$, are
\begin{eqnarray}
  \vec{s}^{(5)}_{1} & = & \vec{s}^{(2)}_{1} + \vec{s}^{(2)}_{2} ,
  \vec{s}^{(5)}_{2} = \vec{s}^{(2)}_{2} + \vec{s}^{(2)}_{3} ,
  \vec{s}^{(5)}_{3} = - \vec{s}^{(2)}_{1} + \vec{s}^{(2)}_{3} , \nonumber \\
  \vec{s}^{(5)}_{4} & = & - \vec{s}^{(5)}_{1} ,
  \vec{s}^{(5)}_{5} = - \vec{s}^{(5)}_{2} ,
  \vec{s}^{(5)}_{6} = - \vec{s}^{(5)}_{3} . \label{s5}
\end{eqnarray}
The $6^{th}$-nearest neighbor vectors, $\vec{s}^{(6)}_{i}$, $i = 1,..., 6$, are
\begin{eqnarray}
  \vec{s}^{(6)}_{1} & = & 2 \vec{s}^{(2)}_{1} ,
  \vec{s}^{(6)}_{2} = 2 \vec{s}^{(2)}_{2} ,
  \vec{s}^{(6)}_{3} = 2 \vec{s}^{(2)}_{3} , \nonumber \\
  \vec{s}^{(6)}_{4} & = & - \vec{s}^{(6)}_{1} ,
  \vec{s}^{(6)}_{5} = - \vec{s}^{(6)}_{2} ,
  \vec{s}^{(6)}_{6} = - \vec{s}^{(6)}_{3} . \label{s6}
\end{eqnarray}
Notice that the number of vectors necessary to reach all the $m^{\textrm{th}}$-near neighbors varies, $i = 1,..., n_m$, but it is always a multiple of 3 (3, 6, 12 for $n_{10}$ etc). A rule is that: When the $\vec{s}^{(m)}_{i}$s vectors connect atoms of the same sublattice (hence contribute to the diagonal terms in the Hamiltonian, see the paper), then $n_m$ is always a multiple of 6, and there is a reflection symmetry between the vectors (e.g., for $m=6$, $\vec{s}^{(6)}_{4} = - \vec{s}^{(6)}_{1}$, $\vec{s}^{(6)}_{5} = - \vec{s}^{(6)}_{2}$, $\vec{s}^{(6)}_{6} = - \vec{s}^{(6)}_{3}$); on the other hand, when the $\vec{s}^{(m)}_{i}$s vectors connect atoms of a different sublattices (hence contribute to the off-diagonal terms in the Hamiltonian, see the paper) then $n_m$ is a multiple of $3$, and there is no reflection symmetry.

Therefore, $V^{(m)}_{\vec{k}}$ is the potential between atoms located at distance $|\vec{s}^{(m)}_i| \equiv |\vec{s}^{(m)}|$, $\forall i = 1,..., n_m$ (in the ideal case of a perfect lattice), while $H^{(\textrm{o})}_{\vec{k}}$ is the Hamiltonian whose eigenfunctions are the orbitals $\varphi^{I} (\vec{r})$ with corresponding eigenvalues $\epsilon^{(0)}$ that is the onsite energy.

Putting together all this, the field Hamiltonian (2) of the main manuscript becomes
\begin{eqnarray}
  {\cal H}_{\vec{k}} & = & \sum_{m} \sum_{i=1}^{n_m} \int d^2 {\vec{r}} {\Large [} a^*_{\vec{k}}  \phi^{* A} (\vec{r})
  (H^{(\textrm{o})}_{\vec{k}} + V^{(m)}_{\vec{k}}) a_{\vec{k}}  \phi^A (\vec{r} + \vec{s}^{(m)}_i) \nonumber \\
  &+& b^*_{\vec{k}}  \phi^{* B} (\vec{r}) (H^{(\textrm{o})}_{\vec{k}} + V^{(m)}_{\vec{k}}) b_{\vec{k}}  \phi^B (\vec{r} + \vec{s}^{(m)}_i) \nonumber \\
  &+& {\Large (} a^*_{\vec{k}}  \phi^{* A} (\vec{r}) (H^{(\textrm{o})}_{\vec{k}} + V^{(m)}_{\vec{k}}) b_{\vec{k}}  \phi^B (\vec{r} + \vec{s}^{(m)}_i) + h.c. {\Large )} {\Large ]}
   \label{fieldHsecond}
\end{eqnarray}
By using that $\phi^{I}_{\vec{k}} (\vec{r}) = e^{i \vec{k} \cdot \vec{r}} \varphi^{I} (\vec{r})$ are Block functions, hence that $\varphi^{I} (\vec{r})$ are periodic, the products in (\ref{fieldHsecond}) all produce phase factors $\phi^{* I} (\vec{r}) \phi^J (\vec{r} + \vec{s}^{(m)}_i) = e^{- i \vec{k} \cdot \vec{r}}  \varphi^{* I} (\vec{r}) e^{i \vec{k} \cdot (\vec{r} + \vec{s}^{(m)}_i)}  \varphi^J (\vec{r} + \vec{s}^{(m)}_i)
= e^{i \vec{k} \cdot \vec{s}^{(m)}_i}  \varphi^{* I} (\vec{r}) \varphi^J (\vec{r} + \vec{s}^{(m)}_i)$. Therefore
\begin{eqnarray}
  {\cal H}_{\vec{k}} & = & \sum_{m \in \textrm{diag}} (\epsilon^{(0)}  \varsigma_m + \eta_m )  {\cal F}_m ({\vec{k}}) (a^*_{\vec{k}} a_{\vec{k}} + b^*_{\vec{k}} b_{\vec{k}}) \nonumber \\
  &+& \left( \sum_{m \in \textrm{off}} (\epsilon^{(0)}  \varsigma_m + \eta_m )  {\cal F}^*_m ({\vec{k}}) a^*_{\vec{k}} b_{\vec{k}} + h.c. \right)   \label{fieldH3rd}
\end{eqnarray}
where
\begin{equation}\label{overlapcoefficients}
  \varsigma_m = \int d^2 \vec{r} \varphi^{* I} (\vec{r}) \varphi^J (\vec{r} + \vec{s}^{(m)}_i) ,
\end{equation}
\begin{equation}\label{hoppingenergies}
  \eta_m = \int d^2 \vec{r} \varphi^{* I} (\vec{r}) V^{(m)} \varphi^J (\vec{r} + \vec{s}^{(m)}_i)
\end{equation}
and ${\cal F}_m ({\vec{k}}) = \sum_{i=1}^{n_m} e^{i {\vec{k}} \cdot {\vec{s}^{(m)}}_i}$, with the $m$s contributing to the diagonal terms when $I = J$, and to the off-diagonal terms when $I \neq J$.

Let us write down here the first six ${\cal F}_m$s
\begin{equation}\label{firstF}
  {\cal F}_1 = \sum_{i= 1}^{3} e^{i {\vec{k}} \cdot {\vec{s}^{(1)}}_i} =
  e^{- i \ell k_y} [1 + 2 e^{i \frac{3}{2} \ell k_y} \cos(\frac{\sqrt{3}}{2} \ell k_x)]
\end{equation}
and
$|{\cal F}_1|^2 = 3 + 2 \sum_{i= 1}^{3} \cos({\vec{k}} \cdot {\vec{s}^{(2)}}_i)$.
\begin{equation}
  {\cal F}_2 = \sum_{i= 1}^{6} e^{i {\vec{k}} \cdot {\vec{s}^{(2)}}_i} =
  2 \sum_{i= 1}^{3} \cos({\vec{k}} \cdot {\vec{s}^{(2)}}_i) = |{\cal F}_1|^2 - 3
\end{equation}
Then, ${\cal F}_3 = \sum_{i= 1}^{3} e^{i {\vec{k}} \cdot {\vec{s}^{(3)}}_i}$, that gives $|{\cal F}_3|^2 = 3 + 2 \sum_{i= 1}^{3} \cos(2 {\vec{k}} \cdot {\vec{s}^{(2)}}_i)$.
\begin{equation}
  {\cal F}_4 = \sum_{i= 1}^{6} e^{i {\vec{k}} \cdot {\vec{s}^{(4)}}_i} =
e^{i\vec{k}\cdot\vec{s}_{1}^{(1)}}\left(e^{-i\vec{k}\cdot\vec{s}_{2}^{(2)}}+e^{-i\vec{k}\cdot\vec{s}_{3}^{(2)}}\right)
+e^{i\vec{k}\cdot\vec{s}_{2}^{(1)}}\left(e^{i\vec{k}\cdot\vec{s}_{2}^{(2)}}+e^{i\vec{k}\cdot\vec{s}_{1}^{(2)}}\right)
+e^{i\vec{k}\cdot\vec{s}_{3}^{(1)}}\left(e^{i\vec{k}\cdot\vec{s}_{3}^{(2)}}+e^{-i\vec{k}\cdot\vec{s}_{1}^{(2)}}\right)
\end{equation}
\begin{eqnarray}
  {\cal F}_5 & = & \sum_{i= 1}^{6} e^{i {\vec{k}} \cdot {\vec{s}^{(5)}}_i} \nonumber \\
& = & 2 (\cos[{\vec{k}} \cdot ({\vec{s}^{(2)}}_1 + {\vec{s}^{(2)}}_2)]     \nonumber \\
& + & \cos[{\vec{k}} \cdot ({\vec{s}^{(2)}}_2 + {\vec{s}^{(2)}}_3)]        \nonumber \\
& + & \cos[{\vec{k}} \cdot ({\vec{s}^{(2)}}_3 - {\vec{s}^{(2)}}_1)])
\end{eqnarray}
\begin{equation}
  {\cal F}_6 = \sum_{i= 1}^{6} e^{i {\vec{k}} \cdot {\vec{s}^{(6)}}_i} =
 2 \sum_{i= 1}^{3} \cos(2 {\vec{k}} \cdot {\vec{s}^{(2)}}_i)= |{\cal F}_3|^2 - 3
\end{equation}
Clearly there is a regularity (pattern) linking the various functions. We shall not investigate this here because it is out of our scope.

Let us now explain out choice of the zero of the energy, and explain, along the way, some more details of the structure of the diagonal and off-diagonal terms.

Define $\widehat{{\cal H}}_{\vec{k}} \equiv \widehat{\tilde{\cal H}}_{\vec{k}} + \epsilon^{(0)} \widehat{{\cal S}}_{\vec{k}}$, where
\begin{equation}\label{overlapmatrix}
 \widehat{{\cal S}}_{\vec{k}} =  \left(
                        \begin{array}{cc}
                          {\cal S}_{\vec{k}}^{\textrm{diag}} &  {\cal S}^{* \textrm{off}}_{\vec{k}} \\
                           {\cal S}^{ \textrm{off}}_{\vec{k}} &  {\cal S}_{\vec{k}}^{\textrm{diag}} \\
                        \end{array}
                      \right)
\end{equation}
is the overlapping matrix, with
\begin{equation}\label{overlapdiag}
  {\cal S}_{\vec{k}}^{\textrm{diag}}  = \sum_{m \in \textrm{diag}} \varsigma_m {\cal F}_m ({\vec{k}}) = 1 + \varsigma_2 {\cal F}_2 ({\vec{k}}) + \varsigma_5 {\cal F}_5 ({\vec{k}}) + \varsigma_6 {\cal F}_6 ({\vec{k}}) + \cdots
\end{equation}
and
\begin{equation}\label{overlapoff}
  {\cal S}_{\vec{k}}^{\textrm{off}} = \sum_{m \in \textrm{off}} \varsigma_m {\cal F}_m ({\vec{k}}) = \varsigma_1 {\cal F}_1 ({\vec{k}}) + \varsigma_3 {\cal F}_3 ({\vec{k}}) + \varsigma_4 {\cal F}_4 ({\vec{k}}) + \cdots
\end{equation}
With these, the Schr\"{o}dinger equation becomes
\begin{equation}\label{Shcroed2}
\psi^\dag_{\vec{k}} \widehat{\tilde{\cal H}}_{\vec{k}} \psi_{\vec{k}} = (E(\vec{k}) - \epsilon^{(0)}) \psi^\dag_{\vec{k}} \widehat{{\cal S}}_{\vec{k}} \psi_{\vec{k}}
\end{equation}
where, as usual, $\psi^\dag_{\vec{k}} \equiv (a^*_{\vec{k}}, b^*_{\vec{k}})$. The energy $\epsilon^{(0)}$ merely shifts the zero of the energy bands, hence we are free to reset the zero of the scale ignoring that term. On the other hand, that zero can still be fixed with the on-site energy included in the Hamiltonian. Thus, by renaming the Hamiltonian as $\widehat{\tilde{\cal H}}_{\vec{k}} \equiv \widehat{{\cal H}}_{\vec{k}}$, we eventually have
\begin{equation}\label{Shcroedtre}
\psi^\dag_{\vec{k}} \widehat{{\cal H}}_{\vec{k}} \psi_{\vec{k}} = E(\vec{k}) \psi^\dag_{\vec{k}} \widehat{{\cal S}}_{\vec{k}} \psi_{\vec{k}}
\end{equation}
where
\begin{equation}\label{firstHwithDirac}
 \widehat{{\cal H}}_{\vec{k}} =
  \left(
                        \begin{array}{cc}
                          h_{\vec{k}}^{\textrm{diag}} &  h^{* \textrm{off}}_{\vec{k}} \\
                           h^{ \textrm{off}}_{\vec{k}} &  h_{\vec{k}}^{\textrm{diag}} \\
                        \end{array}
                      \right)
\end{equation}
with
\begin{equation}\label{hamiltondiag}
 h_{\vec{k}}^{\textrm{diag}} =  \sum_{m \in \textrm{diag}} \eta_m {\cal F}_m ({\vec{k}}) = \eta_0 + \eta_2 {\cal F}_2 ({\vec{k}}) + \eta_5 {\cal F}_5 ({\vec{k}}) + \eta_6 {\cal F}_6 ({\vec{k}}) + \cdots
\end{equation}
\begin{equation}\label{hamiltonoff}
 h^{ \textrm{off}}_{\vec{k}} = \sum_{m \in \textrm{off}} \eta_m {\cal F}_m ({\vec{k}}) = \eta_1 {\cal F}_1 ({\vec{k}}) + \eta_3 {\cal F}_3 ({\vec{k}}) + \eta_4 {\cal F}_4 + \cdots
\end{equation}

\section{Modified commutation relations}
\label{appendix_commutators}

We defined the modified momentum as (see conventions in the main manuscript)
\begin{eqnarray}\label{high-low_trsfn}
\vec{P}&=&\vec{p}+\Delta\vec{p}\;,\nonumber\\
\Delta\vec{p}&=&\frac{\ell}{4}\left(p_{y}^{2}-p_{x}^{2}-\frac{\ell}{2}p_{x}(p^{2}_{x}+p^{2}_{y}),2p_{x}p_{y}-\frac{\ell}{2}p_{y}(p^{2}_{x}+p^{2}_{y})\right) \nonumber \\
&=&\frac{\ell}{4}p^{2}(-\cos2\theta-\frac{\ell}{2}p\cos\theta,\sin2\theta-\frac{\ell}{2}p\sin\theta) \;,
\end{eqnarray}
where the low energy momentum $\vec{p}$ fulfills the standard commutation relation
\begin{equation}\label{standard_commutator}
\left[x_{i},p_{j}\right]=i\hbar\delta_{ij}\;.
\end{equation}
In order to get the commutation relation with the modified momentum, we use the standard commutators (\ref{standard_commutator}) in (\ref{high-low_transformation}), leading to
\begin{eqnarray*}
\left[x,P_{x}\right] & = & i \left(1 - \frac{\ell}{2} P_{x}-\frac{\ell^{2}}{8}(P^{2}_{x}-P^{2}_{y})\right) \; , \nonumber \\
\left[x,P_{y}\right] & = & i \left(\frac{\ell}{2} P_{y}-\frac{\ell^{2}}{4}P_{x}P_{y}\right) \nonumber \\
\left[y,P_{x}\right] & = & i \left(\frac{\ell}{2} P_{y}-\frac{\ell^{2}}{4}P_{x}P_{y}\right)   \; , \\
\left[y,P_{y}\right] & = & i \left(1 + \frac{\ell}{2} P_{x}+\frac{\ell^{2}}{8}(P^{2}_{x}-P^{2}_{y})\right) \;. \nonumber
\end{eqnarray*}
By inverting (13) in the main paper, we get
\begin{equation}\label{P}
{\vec{P}} \equiv {\vec{Q}} (1 + A |{\vec{Q}}|)\;,
\end{equation}
from which
\begin{eqnarray*}
\left[x,Q_{x}\right] & = & i \left(1 - \frac{\ell}{2} Q_{x}-A\left(|\vec{Q}|+\frac{Q_{x}^{2}}{|\vec{Q}|}\right)-\frac{\ell^{2}}{8}(Q^{2}_{x}-Q^{2}_{y})-\frac{\ell}{2}AQ_{x}|\vec{Q}|-A^{2}\left(|\vec{Q}|^{2}+Q_{x}^{2}\right)\right) \; , \nonumber \\
\left[x,Q_{y}\right] & = & i \left( \frac{\ell}{2} Q_{y}-\frac{A}{|\vec{Q}|}Q_{x}Q_{y}-\frac{\ell^{2}}{4}Q_{x}Q_{y}-\frac{\ell}{2}AQ_{x}|\vec{Q}|+\frac{\ell}{2|\vec{Q}|}AQ_{y}\left(Q_{x}^{2}-Q_{y}^{2}\right)-\frac{\ell}{2}A|\vec{Q}|Q_{y}\right) \; , \nonumber \\
\left[y,Q_{x}\right] & = &  i \left( \frac{\ell}{2} Q_{y}-\frac{A}{|\vec{Q}|}Q_{x}Q_{y}-\frac{\ell^{2}}{4}Q_{x}Q_{y}-\frac{\ell}{2}AQ_{x}|\vec{Q}|-\frac{\ell}{2|\vec{Q}|}AQ_{y}Q_{x}^{2}-\frac{\ell}{2}A|\vec{Q}|Q_{y}\right)   \; , \\
\left[y,Q_{y}\right] & = & \left(1 + \frac{\ell}{2} Q_{x}-A\left(|\vec{Q}|+\frac{Q_{y}^{2}}{|\vec{Q}|}\right)+\frac{\ell^{2}}{8}(Q^{2}_{x}-Q^{2}_{y})-\ell \frac{A}{|\vec{Q}|}Q_{x}Q_{y}^{2}-A^{2}\left(|\vec{Q}|^{2}+Q_{y}^{2}\right)\right)\;, \nonumber
\end{eqnarray*}
where $A\equiv \ell{\epsilon'}_1/ \eta_1$.

\bibliography{paper_arXiv_2_biblio}{}
\bibliographystyle{apsrev4-1}

\end{document}